\documentclass{desyproc}

\newcommand{\etal}{{\it et al.}}
 
\begin{document}
%------------------------------------
\title{Tests of Lorentz symmetry}

%for single authors the superscripts are optional
\author{{\slshape Ralf Lehnert}\\[1ex]
Instituto de Ciencias Nucleares,
Universidad Nacional Aut\'onoma de M\'exico\\
A.~Postal 70-543, 04510 M\'exico D.F., Mexico}

% if the proceedings are available online (e.g. at Indico)
% please enter the contribution ID or file_name below for the DOI
%\contribID{32}
%\contribID{lindner\_axel}

% TO THE CONFERENCE EDITORS: 
% please update the following information      
% before sending the template to the authors
% \confID{800}  % if the conference is on Indico uncomment this line
\desyproc{DESY-PROC-2009-05}
\acronym{Patras 2009} % if you want the Acronym in the page footer uncomment this line
%\doi  % if there is an online version we will register DOIs

\maketitle

\begin{abstract}
A number of approaches to fundamental 
physics can lead to the violation of 
Lorentz and CPT symmetry. This talk 
discusses the low-energy phenomenology 
associated with such effects and 
reviews various sample experiments 
within this context.
\end{abstract}

{\em Introduction.}---The Standard Model (SM) and General Relativity (GR) 
provide and excellent phenomenological description of nature.
However, 
from a theoretical viewpoint 
these two theories leave unanswered 
a variety of key 
conceptual 
questions. 
It is therefore believed that the SM and GR 
merge into a single unified theory at high energies
that resolves these issues. 
One possibility for experimental research 
in this field 
is to increase the energy in experiments 
and hope to excite new degrees of freedom, 
which can give insight into such a unified theory.

A complementary experimental approach 
is characterized by tests at comparatively low or moderate energies, 
but with ultra-high precision.
Various efforts along these lines, 
such as searches for axions, 
axion-like particles, 
weakly interacting massive particles,
and weakly interacting sub-eV particles,
have already been discussed at this meeting.
This presentation is focused on
another class of precision experiments, 
namely tests of Lorentz and CPT symmetry.

The special theory of relativity 
and its underlying Lorentz symmetry
have been established over a century ago. 
Since that time, 
Lorentz symmetry has been subjected 
to numerous tests, 
but no credible experimental evidence 
for departures from Lorentz symmetry
has been found. 
In fact, 
special relativity has matured into 
one of the most important cornerstones of physics.
It provides not only the basis for present-day physics,
but it is also the starting point for most 
theoretical approaches to new physics
beyond the SM and GR.

In recent years, 
however, 
it has been realized 
that various of these approaches to new physics 
(although built on Lorentz invariance) 
can accommodate mild, minuscule deviations 
from this symmetry in the ground state~\cite{cpt07}.
Examples of candidate underlying models 
with the possibility of Lorentz violation 
are strings, 
loop quantum gravity,
cosmologically varying scalars,
non-commutative geometry,
and multiverses~\cite{lotsoftheory}.
A further motivation for Lorentz and CPT tests 
is provided by 
the fundamental character of these symmetries: 
they should be backed by experimental evidence 
of steadily increasing quality.

At energy regimes below the Planck scale, 
such departures from Lorentz and CPT symmetry 
can be described in great generality 
by the Standard-Model Extension (SME)~\cite{sme}.
The SME is an effective field theory 
that contains both the usual SM and GR. 
The remaining terms in the SME Lagrangian
control the extent of Lorentz and CPT breakdown;
they are constructed 
to involve all operators for Lorentz and CPT violation 
that are scalars
under coordinate changes.
This broad scope 
guarantees widest applicability:
it eliminates
the association to a particular underlying theory 
and ensures 
that practically all present and near-future experiments 
can be analyzed 
with regards to their potential to measure Lorentz and CPT symmetry. 
Numerous investigations 
have been performed within the SME~\cite{theory},
which confirm its sound theoretical basis. 
The SME has become the standard framework 
for the identification and analysis 
of a wide variety of experimental studies~\cite{cpt07,kr}.
For instance, 
the SME predicts modifications in one-particle dispersion relations~\cite{thres}, 
which in turn could lead to vacuum Cherenkov radiation~\cite{cher}, 
corrections to the usual Compton edge,
or changes in neutral-meson oscillations. 
The absence of these effects
leads to tight constraints on Lorentz violation~\cite{hohen08,collider}. 
For other limits on violation, 
see, e.g., Refs.~\cite{CFJ,gamma,neutron}.

{\em The SME test framework.}---A test framework 
that allows for departures from Lorentz and CPT symmetry
is useful for the identification and analysis 
of suitable experiments. 
Establishing such a test model 
requires some preliminary thoughts.
One issue is the multitude of approaches to underlying physics 
that can lead to Lorentz and CPT violation: 
there is presently no single realistic candidate fundamental theory 
whose low-energy limit can serve as the test framework. 
A related difficulty is the fact that 
for some theories beyond the SM and GR 
the low-energy limit is unknown or not unique. 
As a consequence, 
the test framework will be constructed by hand 
with the objective of relative independence 
of the details of the underlying physics. 

The first step is to determine 
how Lorentz and CPT breakdown 
can be implemented into the test framework. 
One possibility 
that has proved to be the most general and useful 
is the inclusion of preferred directions 
modeled by background vectors and tensors 
while leaving the Lorentzian structure of spacetime unaffected. 
This idea is compatible with the fact
that most candidate underlying models 
take Lorentz symmetry as a key basic ingredient. 
Once the model's dynamics is taken as Lorentz symmetric, 
symmetry breaking can essentially only arise
along the lines of a Lorentz-violating ground state.
Moreover, 
this implementation of deviations from Lorentz and CPT symmetry
can maintain coordinate independence, 
a principle more fundamental than Lorentz symmetry. 
An immediate consequence is 
that different inertial coordinate systems 
are still connected via the usual Lorentz transformations. 
Violations of the symmetry become apparent only
through the physical transformations:
boosts and rotations of the experimental set-up;
the background vectors and tensors are 
outside of experimental control and remain fixed
under such physical transformations.

The springboard for the construction of the SME 
is the SM Lagrangian ${\cal L}_{\rm SM}$ 
and the Einstein--Hilbert Lagrangian ${\cal L}_{\rm EH}$,
which essentially contain the entire body of present-day physics. 
This guarantees that 
departures from Lorentz and CPT symmetry 
in all known physical systems 
can be described. 
The small Lorentz- and CPT-violating corrections $\delta{\cal L}_{\rm LIV}$ 
are formed by contracting the background vectors and tensors 
with ordinary SM and gravitational fields 
to yield scalars under coordinate changes:
\begin{equation}
\label{smelagr}
{\cal L}_{\rm SME}={\cal L}_{\rm SM}+{\cal L}_{\rm EH}+\delta{\cal L}_{\rm LIV}\;.\end{equation}
Examples of terms present in the Minkowski-spacetime limit of $\delta{\cal L}_{\rm LIV}$ are
\begin{equation}
\label{sampleterms}
\delta{\cal L}_{\rm LIV}\supset b_\mu\,\overline{\psi}\gamma^{\mu}\gamma_5\psi,\; 
(r_\mu\,\overline{\psi}\gamma^{\mu}\gamma_5\psi)^2,\; 
(k_F)^{\alpha\beta\gamma\delta}F_{\alpha\beta}F_{\gamma\delta},\;
(k_{AF})^{\alpha}A^{\beta}\tilde{F}_{\alpha\beta},\;
\ldots
\;.
\end{equation}
Here, $\psi$, $A$, and $F$ 
are a conventional spinor field,
a conventional gauge potential,
and a conventional gauge field strength, respectively. 
The non-dynamical 
$b_\mu$, $r_\mu$, $(k_F)^{\alpha\beta\gamma\delta}$, and $(k_{AF})^{\alpha}$ 
are minute Lorentz-violating background vectors and tensors 
assumed to be generated by a candidate fundamental theory. 
Experimental tests seek to bound or measure these vectors and tensors.
We finally mention 
that the minimal SME (mSME) is restricted by further physical requirements, 
such as translational invariance, the usual gauge symmetries, 
and power-counting renormalizability. 
For example, the mSME does not contain the $r_\mu$ term 
present in the above expression~(\ref{sampleterms}).

{\em Lorentz violation via varying scalars.}---In 
the construction of the SME, 
we have included 
the external non-dynamical background vectors and tensors 
that select preferred directions 
by hand
without reference to underlying physics.
A natural question to ask is
whether such Lorentz-violating preferred directions
can really be generated 
by candidate fundamental theories. 
We will briefly discuss one example 
illustrating 
that this is indeed the case: 
varying scalars. 

Many theoretical approaches to underlying physics
predict novel scalar fields. 
In fact, 
certain cosmological observations, 
such as the flatness and the accelerated expansion of the universe, 
can be explained by invoking new scalars.
In such cosmological contexts, 
scalar fields can acquire nonzero expectation values 
with time dependencies driven 
by the evolution of the scale factor. 
As one example, 
we may consider $N=4$ supergravity in four spacetime dimensions,
which contains novel axion $a$ and dilation $b$ fields
coupled via a function $f(a,b)$ to the electromagnetic field strength.
In a simple cosmological model
one can determine the evolution of $a$ and $b$ 
with the comoving time $t$. 
One of the couplings to electrodynamics 
then generates the effective Lagrangian term $f(t)F\tilde{F}$.
In a local, experimental setting, 
such a term would indeed be perceived as a varying coupling---in this case, 
as a time-dependent $\theta$ angle.

A spacetime-dependent scalar, 
regardless of the mechanism that causes the variation, 
normally implies the breakdown of spacetime-translation invariance. 
But also Lorentz symmetry is typically violated in such situations
because the gradient of the scalar selects a preferred direction. 
At the formal level,  
this fact is intuitively reasonable: 
the definition of the Lorentz-transformation generators 
contains the energy--momentum tensor, 
which is now no longer conserved. 
Thus, 
the usual time-independent boost and rotation generators 
no longer exist.
To see this explicitly in our toy supergravity model,
we can perform an integration by parts in the action: 
\begin{equation}
\label{CST}
f(t)F\tilde{F}\to-2\left(\partial^\alpha f \right)A^\beta\tilde{F}_{\alpha\beta}\; .
\end{equation}
The cosmological background $f(t)$ is essentially outside of experimental control
for the purposes of local measurements, 
so $\partial^\alpha f$ can be taken as non-dynamical.
If we identify $-2\left(\partial^\alpha f \right)$ 
with $(k_{AF})^{\alpha}$ in Eq.~(\ref{sampleterms}), 
we explicitly see how this Lorentz- and CPT-violating 
Chern--Simons-type correction~\cite{CFJ}
can be generated by underlying physics.

{\em Experimental tests.}---Since Lorentz symmetry underpins
many areas and concepts in physics, 
it can be tested in a multitude of physical systems. 
The tests with the best potential for highest sensitivity 
can be identified and analyzed with the SME. 
We briefly discuss three sample tests.

The first example concerns an astrophysical search 
for the Cherns--Simons-type term~(\ref{CST}) 
mentioned earlier. 
A theoretical analysis of this term reveals 
that it causes birefringence. 
Even the smallest birefringence effects 
would accumulate for light 
that has traveled a sufficiently large distance. 
It is therefore unsurprising 
that the best experimental constraints 
on this particular type of Lorentz- and CPT-violation
have been obtained from observations
of cosmological sources.
One predicted effect would be following.
Suppose an astrophysical source is emitting 
flashes of light containing all polarizations. 
En route to Earth, 
such a pulse would separate 
because one of its two components travels faster 
than the other
due to birefringence. 
A somewhat more sophisticated and sensitive approach 
is to observe a cosmological object 
known to emit a spectrum of mostly polarized light 
and measure the polarization of this light 
as a function of its wavelength at Earth.
For birefringence due to a Chern--Simons-type interaction~(\ref{CST}), 
this function should display a predicted characteristic. 
Such analyses have indeed been performed, 
and no such characteristic was found. 
This implies the bound $(k_{AF})^{\alpha}<10^{-43}\,$GeV~\cite{CFJ,gamma}.

The second sample Lorentz test
involves (anti)protons in Penning traps.
The basic idea is as follows. 
The Lorentz-violating preferred background directions 
act in many respects just like external fields. 
In conventional physics, 
such external fields can cause level shifts in bound systems 
like atoms
(e.g., the Zeeman and Stark effects).
Calculations within the SME reveal
that Lorentz and CPT breakbown 
would cause similar level shifts for charges 
in Penning traps, for example.
More precisely, 
the anomaly transitions would acquire opposite corrections 
for protons and their antiparticles.
This fact can be employed to 
extract clean experimental limits 
on the $b^\mu$ coefficient (see expession~(\ref{sampleterms})) for the proton
with a sensitivity of about $10^{-24}\,$GeV~\cite{penning}.

In the experimental investigations discussed above, 
gravitational effects could be neglected 
and the flat-spacetime limit of the SME was considered. 
However, 
tests involving gravity have recently been one focus of attention~\cite{grav,kt08}.
In particular, 
antimatter, 
such as antihydrogen, 
offers the possibility of testing Lorentz and CPT symmetry 
in the SME's gravity sector. 
For instance, 
the acceleration of antihydrogen 
in the Earth's gravitational field could be investigated~\cite{kt08}. 
We also note that in gravitational contexts,
various SME coefficients 
that are inaccessible in the flat-spacetime limit
now become measurable~\cite{kt08}.

{\em Acknowledgments.}---The author wishes to thank J\"org J\"ackel 
for the invitation to this stimulating meeting. 
This work is supported in part
by CONACyT under Grant No.~55310.

% ****************************************************************************
% BIBLIOGRAPHY AREA
% ****************************************************************************

\begin{footnotesize}
% IF YOU DO NOT USE BIBTEX, USE THE FOLLOWING SAMPLE SCHEME FOR THE REFERENCES
% ----------------------------------------------------------------------------

% ----------------------------------------------------------------------------

% IF YOU USE BIBTEX,
% - DELETE THE TEXT BETWEEN THE TWO ABOVE DASHED LINES
% - UNCOMMENT THE NEXT TWO LINES AND REPLACE 'Name_Of_Your_BibFile'

%\bibliographystyle{unsrt}
%\bibliography{Name_Of_Your_BibFile}

\begin{thebibliography}{99}
%------- replace following references ;-)
\bibitem{cpt07}
For recent reviews 
see, e.g.,
V.A.\ Kosteleck\'y, ed.,
{\it CPT and Lorentz Symmetry I-IV},
World Scientific, Singapore, 1999-2008;
R.\ Bluhm,
Lect.\ Notes Phys.\ {\bf 702}, 191 (2006)
[hep-ph/0506054].
%%CITATION = LNPHA,702,191;%%

\bibitem{lotsoftheory}
See, e.g., V.A.\ Kosteleck\'y and S.\ Samuel,
Phys.\ Rev.\ D {\bf 39}, 683 (1989);
%%CITATION = PHRVA,D39,683;%%
V.A.\ Kosteleck\'y and R.\ Potting,
Nucl.\ Phys.\ B {\bf 359}, 545 (1991);
%%CITATION = NUPHA,B359,545;%%
J.\ Alfaro, H.A.\ Morales-T\'ecotl, and L.F.\ Urrutia,
Phys.\ Rev.\ Lett.\  {\bf 84}, 2318 (2000); 
%%CITATION = PRLTA,84,2318;%%
S.M.\ Carroll \etal,
Phys.\ Rev.\ Lett.\ {\bf 87}, 141601 (2001);
%%CITATION = PRLTA,87,141601;%%
V.A.\ Kosteleck\'y, R.\ Lehnert, and M.J.\ Perry,
Phys.\ Rev.\ D {\bf 68}, 123511 (2003);
%%CITATION = PHRVA,D68,123511;%%
O.\ Bertolami \etal,
Phys.\ Rev.\ D {\bf 69}, 083513 (2004);
%%CITATION = PHRVA,D69,083513;%%
J.D.\ Bjorken,
Phys.\ Rev.\ D {\bf 67}, 043508 (2003);
%%CITATION = PHRVA,D67,043508;%%
N.\ Arkani-Hamed \etal, 
JHEP {\bf 0701}, 036 (2007).
%%CITATION = JHEPA,0701,036;%%

\bibitem{sme}
D.\ Colladay and V.A.\ Kosteleck\'y,
Phys.\ Rev.\ D {\bf 55}, 6760 (1997);
%%CITATION = PHRVA,D55,6760;%%
Phys.\ Rev.\ D {\bf 58}, 116002 (1998); 
%%CITATION = PHRVA,D58,116002;%%
V.A.\ Kosteleck\'y and R.~Lehnert,
Phys.\ Rev.\  D {\bf 63}, 065008 (2001);
%%CITATION = PHRVA,D63,065008;%%
V.A.\ Kosteleck\'y,
Phys.\ Rev.\ D {\bf 69}, 105009 (2004);
%%CITATION = PHRVA,D69,105009;%%
V.A.~Kosteleck\'y and M.~Mewes,
Phys.\ Rev.\  D {\bf 80}, 015020 (2009).
%%CITATION = PHRVA,D80,015020;%%


\bibitem{theory} 
See, e.g., 
R.~Jackiw and V.A.~Kosteleck\'y,
Phys.\ Rev.\ Lett.\  {\bf 82}, 3572 (1999);
%%CITATION = PRLTA,82,3572;%%
V.A.~Kosteleck\'y \etal,
Phys.\ Rev.\  D {\bf 65}, 056006 (2002);
%%CITATION = PHRVA,D65,056006;%%
B.~Altschul and V.A.~Kosteleck\'y,
Phys.\ Lett.\  B {\bf 628}, 106 (2005);
%%CITATION = PHLTA,B628,106;%%
R.~Lehnert, 
J.\ Math.\ Phys.\  {\bf 45}, 3399 (2004);
%%CITATION = JMAPA,45,3399;%%
Phys.\ Rev.\  D {\bf 74}, 125001 (2006);
%%CITATION = PHRVA,D74,125001;%%
arXiv:0711.4851 [hep-th];
%%CITATION = ARXIV:0711.4851;%%
A.J.~Hariton and R.~Lehnert,
Phys.\ Lett.\  A {\bf 367}, 11 (2007);
%%CITATION = PHLTA,A367,11;%%
J.~Alfaro \etal,  
Int.\ J.\ Mod.\ Phys.\  A {\bf 25}, 3271 (2010);
%%CITATION = IMPAE,A25,3271;%%
D.~Colladay, P.~McDonald, and D.~Mullins,
J.\ Phys.\ A  {\bf 43}, 275202 (2010).
%%CITATION = JPAGB,A43,275202;%%

\bibitem{kr}
V.A.~Kosteleck\'y and N.~Russell,
arXiv:0801.0287v3 [hep-ph].
%%CITATION = ARXIV:0801.0287;%%

\bibitem{thres}
R.~Lehnert,
Phys.\ Rev.\  D {\bf 68}, 085003 (2003).
%%CITATION = PHRVA,D68,085003;%%

\bibitem{cher}
R.~Lehnert and R.~Potting, 
Phys.\ Rev.\ Lett.\  {\bf 93}, 110402 (2004);
%%CITATION = PRLTA,93,110402;%%
Phys.\ Rev.\  D {\bf 70}, 125010 (2004);
%%CITATION = PHRVA,D70,125010;%%
K.G.~Zloshchastiev, arXiv:1003.0657 [hep-th].
%%CITATION = ARXIV:1003.0657;%%

\bibitem{hohen08}
M.A.~Hohensee \etal,  
Phys.\ Rev.\ Lett.\  {\bf 102}, 170402 (2009);
%%CITATION = ARXIV:0904.2031;%%
Phys.\ Rev.\  D {\bf 80}, 036010 (2009).
%%CITATION = PHRVA,D80,036010;%%

\bibitem{collider}
J.-P.~Bocquet {\it et al.},
Phys.\ Rev.\ Lett.\  {\bf 104}, 241601 (2010);
%%CITATION = PRLTA,104,241601;%%
G.~Amelino-Camelia {\it et al.},
Eur.\ Phys.\ J.\  C {\bf 68}, 619 (2010).
%%CITATION = EPHJA,C68,619;%%

\bibitem{CFJ}
S.M.\ Carroll, G.B.\ Field, and R.\ Jackiw, 
Phys.\ Rev.\ D {\bf 41}, 1231 (1990). 
%%CITATION = PHRVA,D41,1231;%% 

\bibitem{gamma}
V.A.~Kosteleck\'y and M.~Mewes,
Phys.\ Rev.\ Lett.\  {\bf 99}, 011601 (2007); 
%%CITATION = PRLTA,99,011601;%%
M.~Mewes,
Phys.\ Rev.\  D {\bf 78}, 096008 (2008).
%%CITATION = PHRVA,D78,096008;%%

\bibitem{neutron}
I.~Altarev {\it et al.},
Phys.\ Rev.\ Lett.\  {\bf 103}, 081602 (2009);
%%CITATION = PRLTA,103,081602;%%
arXiv:1006.4967 [nucl-ex].
%%CITATION = ARXIV:1006.4967;%%

\bibitem{penning}
R.~Bluhm, V.A.~Kosteleck\'y, and N.~Russell,
Phys.\ Rev.\  D {\bf 57}, 3932 (1998).
%%CITATION = PHRVA,D57,3932;%%

\bibitem{grav}
Q.G.~Bailey and V.A.~Kosteleck\'y,
Phys.\ Rev.\  D {\bf 74}, 045001 (2006);
%%CITATION = PHRVA,D74,045001;%%
H.~M\"uller \etal,
Phys.\ Rev.\ Lett.\  {\bf 100}, 031101 (2008);
%%CITATION = PRLTA,100,031101;%%
J.B.R.~Battat, J.F.~Chandler, and C.W.~Stubbs,
Phys.\ Rev.\ Lett.\  {\bf 99}, 241103 (2007);
%%CITATION = PRLTA,99,241103;%%
Q.G.~Bailey,
Phys.\ Rev.\  D {\bf 80}, 044004 (2009).
%%CITATION = PHRVA,D80,044004;%%

\bibitem{kt08}
V.A.~Kosteleck\'y and J.~Tasson,
Phys.\ Rev.\ Lett.\  {\bf 102}, 010402 (2009).
%%CITATION = PRLTA,102,010402;%%

\end{thebibliography}
% example of Name_Of_Your_BibFile.bib
% @Article{Turcato:2006ch,
%      author    = "Turcato, M.",
%  collaboration = "ZEUS and H1",
%      title     = "Lepton flavour violation and charmonium physics at HERA",
%      journal   = "Nucl. Phys. Proc. Suppl.",
%      volume    = "162",
%      year      = "2006", 
%      pages     = "283-287",
%      SLACcitation  = "%%CITATION = NUPHZ,162,283;%%"
% }
% 
% @Unpublished{Gogitidze:2007du,
%      author    = "Gogitidze, N.",
%  collaboration = "H1", 
%      title     = "Prompt photons and particle momentum distributions at
%                   HERA", 
%      year      = "2007",
%      note    = "hep-ex/0701033",
%      SLACcitation  = "%%CITATION = HEP-EX 0701033;%%"
% }

\end{footnotesize}

% ****************************************************************************
% END OF BIBLIOGRAPHY AREA
% ****************************************************************************

\end{document}